# Broadband THz spectroscopy system beyond 25 THz using BNA crystals and a tunable single-ring-fiber compressor

Wei Cui,[1,*] Aswin Vishnuradhan,[1] Markus Lippl,[2,3] Eeswar Kumar Yalavarthi,[1] Angela Gamouras,[1,4] Nicolas Y. Joly,[2,3,5] and Jean-Michel Ménard[1,4,*]

[1]Department of Physics, University of Ottawa, 25 Templeton Street, Ottawa, Ontario, Canada, K1N 6N5
[2]Max Planck Institute for the Science of Light, Staudtstr. 2, Erlangen, Bavaria, Germany, 91058
[3]Department of Physics, University of Erlangen-Nürnberg, Staudtstr. 2, Erlangen, Bavaria, Germany, 91058
[4]National Research Council Canada, 1200 Montreal Road, Ottawa, Ontario, Canada, K1A 0R6
[5]Integrated center for nanostructured films, Cauerstr. 3, Erlangen, Bavaria, Germany, 91058

**ABSTRACT**. We present a terahertz (THz) time-domain spectroscopy (THz-TDS) system which accesses a broadband spectrum, efficiently covering the so-called "new THz gap" between 5 and 15 THz and extending beyond 25 THz. The system exploits nonlinear interactions within the organic crystal BNA (N-benzyl-2-methyl-4-nitroaniline) to generate and detect THz radiation upon excitation by a near-infrared (NIR) pulse centered at 1.03 μm. To enable broadband THz spectral monitoring, the NIR pulse from a Yb-based solid-state laser undergoes spectral broadening in a gas-filled single-ring hollow-core photonic-crystal fiber, followed by a pulse compression to achieve durations as short as 31 fs. This approach paves the way for broadband spectroscopy in hard-to-access THz regions using widely available near-infrared ultrafast sources.

## I. INTRODUCTION.

Terahertz time-domain spectroscopy (THz-TDS) is a non-invasive technique capable of phase-sensitive measurements, allowing direct access to the complex dielectric properties of materials [1]. The technique has found widespread use in scientific research to monitor a vast range of fundamental excitations in fascinating solid-state systems such as superconductors [2–4], topological insulators [5,6], Dirac and Weyl semimetals [7,8], heavy fermion systems [9,10], and quantum spin liquids [11,12]. Furthermore, due to its capacity to distinguish unique spectral signatures of organic and inorganic materials, the technique holds significant potential for industrial quality monitoring. Expanding the operational bandwidth of THz-TDS is essential for accessing more information as it not only advances our understanding of the underlying physics in emerging materials but also enables new applications in sensing and imaging [13–15].

The development of broadband THz-TDS systems has been in part limited by the availability of nonlinear devices allowing efficient THz generation and detection at frequencies between 5 and 15 THz [16,17]. Photoconductive methods for THz emission and detection are extensively employed because of their straightforward design and compact structure. However, phonon absorption and electronic dynamics usually prevent these devices to efficiently reach frequencies above 10 THz [18]. Other systems using optical rectification (OR) for THz generation and electro-optic sampling (EOS) for detection, rely on noncentrosymmetric materials featuring a strong second-order ($\chi(2)$) nonlinearity, favorable phase-matching conditions, and low absorption in both the NIR and THz spectral windows. Finding a material with a relatively low absorption over a large THz window is particularly difficult to satisfy, as most materials exhibit strong absorption between 5 and 15 THz due to optical phonons [16,17]. One way to circumvent these material limitations is by using two-color laser-induced plasma generation in gas [15,19], which must be combined with air-based coherent detection [20]. This scheme, however, requires a mJ-pulse-energy laser source and a relatively complex optical configuration. Spintronic emitters have also emerged as capable sources to generate broadband THz radiation without being limited by phase-matching constraints [21]. However, their optical-to-terahertz conversion efficiency has remained so far lower than many optical rectification techniques [22]. As another alternative, broadband THz pulses can be generated with Raman-resonance-enhanced four-wave mixing in diamond. However, this technique requires tightly synchronized multi-pulse excitation, precise angular alignment for phase matching, and high peak intensities, making it experimentally more complex than conventional methods [23].

The emergence of organic crystals offers new possibilities for THz-TDS. They exhibit a high $\chi(2)$ nonlinearity, satisfy phase-matching conditions for three-wave mixing processes involving NIR wavelengths, and feature low THz absorption thanks to weaker phonon resonances compared with semiconductors. Most organic materials studied to date in the context of THz generation, such as DAST, OH1, and DSTMS, satisfy phase-matching conditions near 1550 nm [24]. At 1030 nm, the typical wavelength of Yb-based ultrafast lasers, DAST and OH1 no longer meet favorable phase-matching conditions while DSTMS produces a spectrum that peaks near 2 THz and extends to about 8 THz [25,26]. In contrast, BNA (N-benzyl-2-methyl-4-nitroaniline) is a particularly well suited for broadband THz spectroscopy when pumped at 1030 nm, supporting a significantly wider generation bandwidth, while showing relatively weak resonant absorption [27–29]. Previous work notably demonstrated high-field THz emission from BNA, reaching peak electric fields of 1 MV/cm [30] and average THz powers of 5.6 mW [27,31,32]. However, the usable THz bandwidth in these systems has remained limited by the nonlinear detection scheme relying on semiconductor crystals, such as GaP. A potential solution toward a broader detected spectrum is to employ organic crystals as EO detectors as well, since they can in principle support wide THz bandwidths [33–36]. In practice, however, such implementations are challenging because the strong birefringence of organic crystals complicates polarization-sensitive detection schemes. Among recent strategies, THz-induced lensing (TIL)–based EOS has been shown to mitigate birefringence-related challenges in strongly anisotropic organic crystals, but this improvement comes with a greater demand for precise optical alignment [33,35]. Kuroyanagi et al. showed that conventional EOS can be implemented using BNA for both generation and detection at 815 nm, achieving a bandwidth of 8 THz [34]. Most recently, Mansourzadeh et al. demonstrated THz generation and conventional EOS detection with MNA, reaching an ~9 THz bandwidth [36]. Collectively, these works illustrate both the promise and the limitations of organic-crystal THz systems, particularly in scaling the detection bandwidth while maintaining high sensitivity at elevated THz frequencies.

Another significant challenge in developing broadband THz-TDS systems is the need for a high-cost ultrafast optical source with a sufficiently short pulse duration and broad spectral bandwidth to enable THz generation and detection at high frequencies. For example, most systems based on commercially available Yb:KGW NIR sources delivering >100 fs pulses centered at 1030 nm typically offer bandwidths of 4 THz or less. To extend the accessible THz window beyond this limit, pulse compressing methods such as multi-pass cell [37], gas-filled hollow-core fibers (HCFs) [38] or gas-filled hollow-core photonic crystal fibers (HC-PCFs) have been demonstrated [39]. Spectrally broadened 1030 nm lasers generated using these techniques have been applied to efficiently produce broadband THz radiation in organic crystals [26,27,30,36]. The main advantage of the HC-PCFs is the ability to control linear dispersion and nonlinearity by adjusting the gas pressure and gas composition within the hollow core. This approach has been successfully demonstrated using an argon-filled kagomé HC-PCF [40], doubling the bandwidth of a THz-TDS system based on GaP crystals for THz generation and detection. Replacing GaP with GaSe further extends the accessible THz spectral range, enabling tunable multi-THz emission between 10 and 18 THz [41]. Recently developed single-ring HC-PCFs retain the key advantages of kagomé HC-PCFs, such as pressure-tunable dispersion and nonlinearity, while avoiding the unpredictable loss bands due to the complex structure and can therefore offer lower loss while being easier to fabricate [42,43].

In this work, we present an ultra-broadband THz-TDS system that uses BNA crystals for both THz generation and detection, combined with a single-ring HC-PCFs to compress the NIR pulses from a commercial Yb:KGW ultrafast amplifier. Yb-based amplifiers offer excellent power scalability and long-term stability [44], qualities that have driven their widespread use in high-duty-cycle ultrafast spectroscopy and nonlinear THz experiments. Unlike recent studies, which focused primarily on THz generation in BNA crystals [27,30], our work demonstrates a BNA-based THz-TDS platform that integrates BNA for both THz generation and electro-optic detection while being pumped by a compressed Yb:KGW laser source. Importantly, our implementation relies on single-ring HC-PCF based spectral broadening followed by chirped-mirror compression of the NIR pulses, enabling broadband generation and detection. Our THz-TDS provides access to the region between 0.3 and 25.2 THz, with a substantial spectral coverage across the “new THz gap” between 5 and 15 THz. This broadband capability provides a powerful platform for probing phonons and polaritons in two-dimensional and semiconductor systems [45–47], as well as emergent excitations in quantum materials [48].

*Contact author: wcui2@uottawa.ca
*Contact author: jean-michel.menard@uottawa.ca

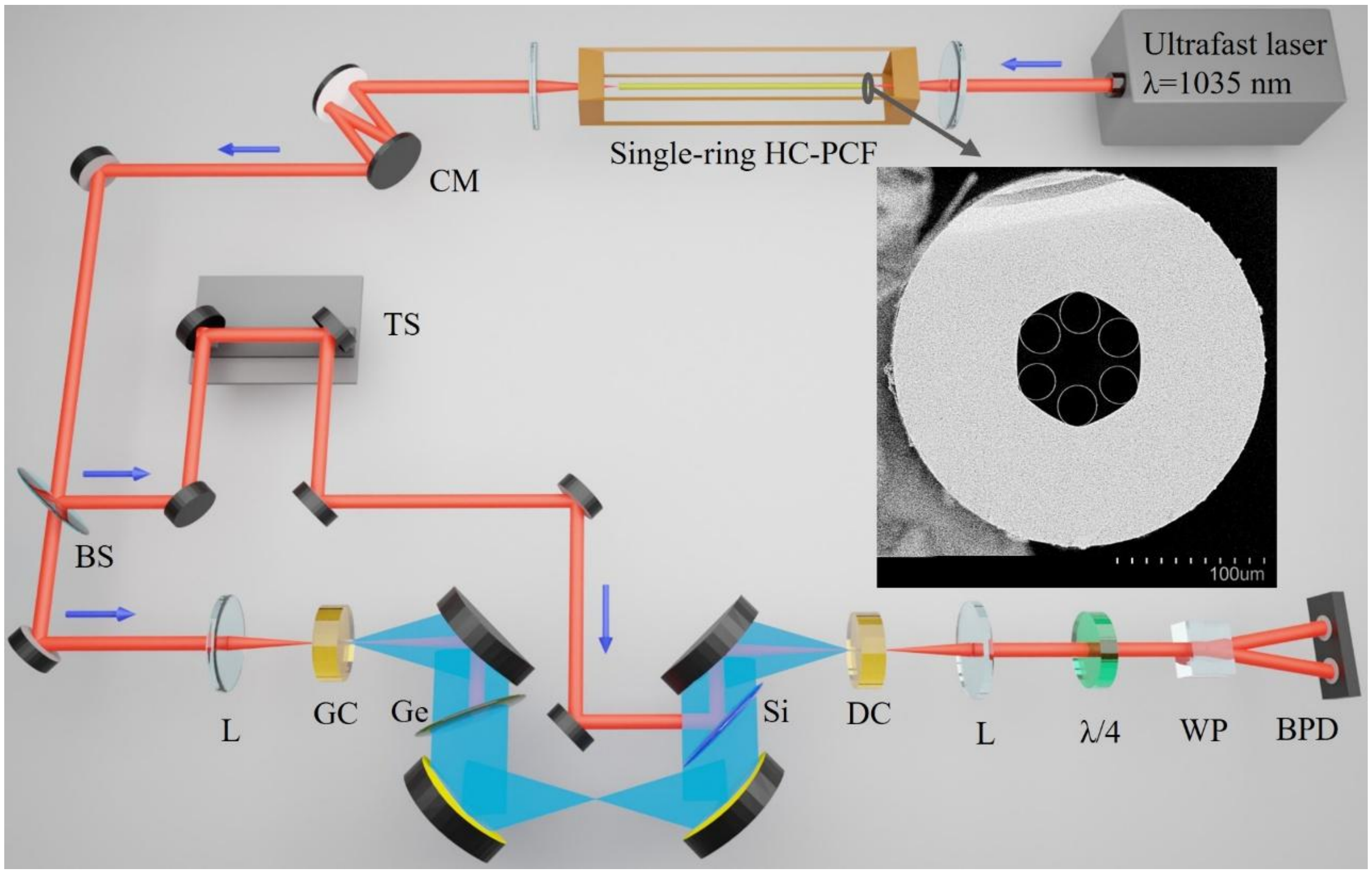

FIG 1. Schematic of the experimental setup. NIR pulses from the ultrafast laser are launched into a single-ring HC-PCF filled with argon gas at 15 bar. A standard THz-TDS system [1,40] is then used to generate and detect THz pulses. The system operates in a dry-air purged environment. CM: chirped mirror; BS: beam splitter; TS: translational stage; L: lens; BNA: N-benzyl-2-methyl-4-nitroaniline generation/detection crystal; Ge: Germanium wafer; Si: Silicon wafer; λ/4: quarter-wave plate; WP: Wollaston prism; BPD: balanced photodetectors.

## II. EXPERIMENTS

Figure 1 shows a schematic of the experimental apparatus. A single-ring HC-PCF is placed before a standard THz-TDS configuration relying on organic BNA crystals for THz generation and detection [40]. The system is driven by a commercial Yb:KGW regenerative amplifier system generating 180 fs pulses centered at 1035 nm and operating at a repetition rate of 1.1 MHz. The NIR pulses are launched into a single-ring HC-PCF with an 85% coupling efficiency. The scanning electron micrograph of the transverse structure of the single-ring HC-PCF is shown in the inset of Fig. 1. The HC-PCF is 55 cm long, with a core diameter D ~ 37 µm, a hollow capillary diameter d ~ 28 µm and a capillary wall thickness of 300 nm. The HC-PCF is enclosed in a gas cell filled with argon gas at 15 bar, which serves as nonlinear medium for spectral broadening. At this pressure the fiber exhibits anomalous dispersion over the used spectral range (zero dispersion lies at 850 nm), therefore the fiber length had to be chosen such that one can only observe self-phase modulation (SPM) and no soliton fission can occur. The length scale for soliton fission to take place is estimated to be 75 cm [49] at the highest energy available at the given laser repetition rate (considering the coupling efficiency), in order to have sufficient safety margin a fiber length of 55 cm is chosen. This limitation means that the negative dispersion of the fiber is not sufficient to yield a transform-limited pulse. To compensate for the remaining positive chirp a pair of chirped mirrors providing a dispersion of -250 fs2 per reflection is placed after the fiber output. The number of chirped-mirror bounces applied to the NIR pulse after the single-ring fiber is determined by continuously varying the number of bounces from 3 to 16 and using intensity autocorrelation measurements to identify the shortest pulse duration. The beam is then guided into a THz-TDS system [40]. In such a system, the beam is split into two arms by a beam splitter (BS). The transmitted beam is focused onto a 400 µm-thick BNA crystal attached to a 500 µm-thick z-cut sapphire to generate THz through optical rectification. The NIR generation pulse incident on the generation crystal has a spot size of 24.8 μm (1/e2 diameter). The excitation beam is attenuated to a pulse energy below 0.15 μJ to prevent damaging the BNA crystal. Another beam, reflected by the BS, is used for THz detection using the EOS technique. The EOS crystal is a free-standing 300 μm-thick BNA crystal. The NIR gating pulse incident on the detection crystal has a spot size of 8.2 μm (1/e2 diameter). A motorized linear translation stage is used to sample the terahertz waveform point by point with a step size of Δx = 1.5 μm, corresponding to a temporal increment of Δt = 10 fs. A lock-in amplifier with a time constant of 100 ms and an additional waiting time of 500 ms at each step enabled sensitive acquisition of 500 temporal points within approximately 8 min. The waiting time dramatically limits the acquisition speed but is especially important here, since the stop-and-go motion of the translation stage induces beam-pointing instabilities. Techniques demonstrated in the literature could be implemented to address this issue, notably asynchronous optical sampling (ASOPS) [50] based on two synchronized ultrafast lasers, as well as other fast-scanning approaches [51,52].

*Contact author: wcui2@uottawa.ca
*Contact author: jean-michel.menard@uottawa.ca

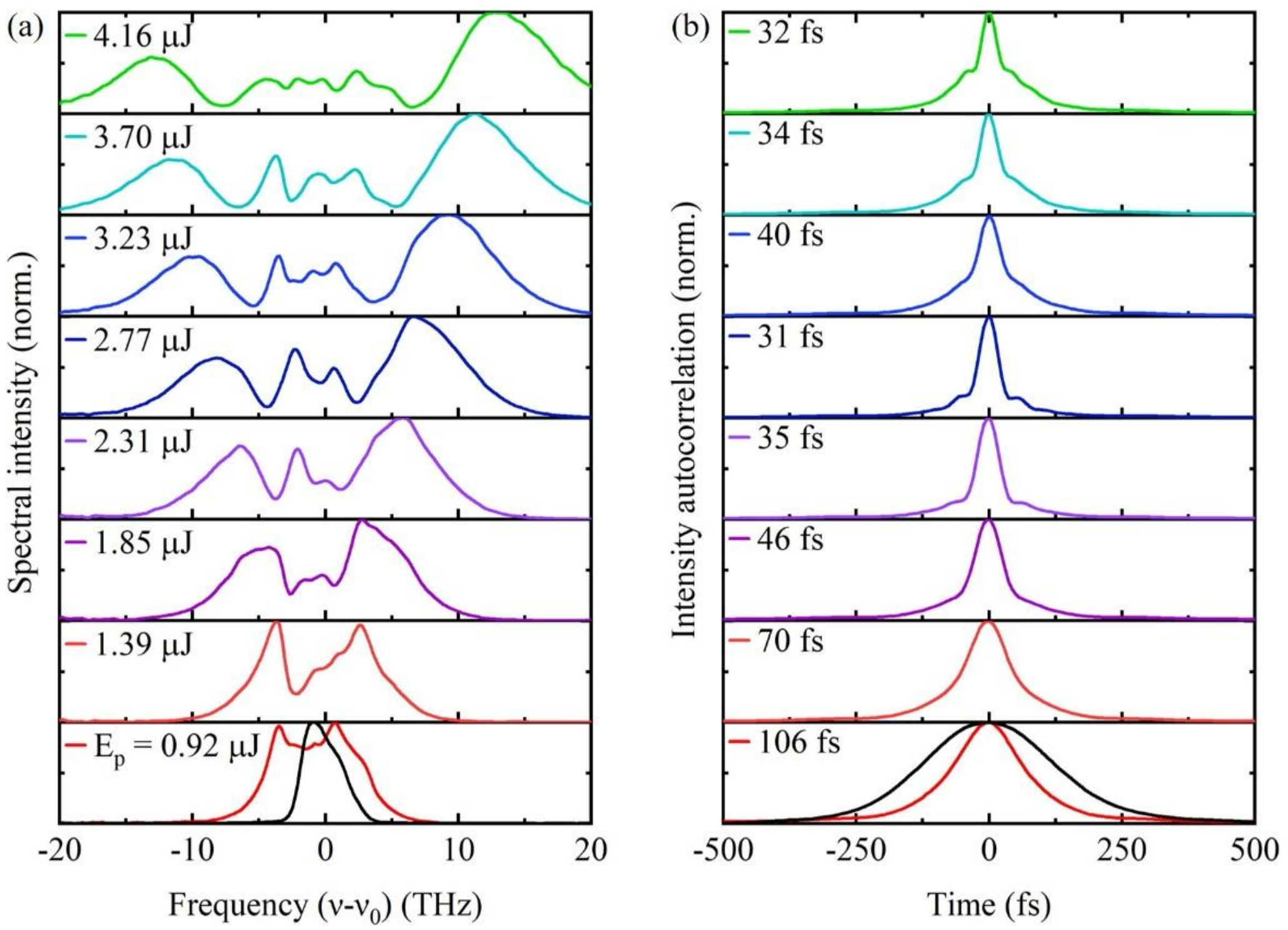

FIG 2. (a) Measured NIR spectra after the single-ring HC-PCF with different input pulse energies $E_p$. (b) Measured intensity autocorrelation traces of the NIR pulses shown in (a) after reflecting off a pair of CMs. The FWHM of the original NIR pulse obtained after deconvolution (assuming a Gaussian pulse shape) is shown in the top-left corner. For $E_p$ going up to 2.77 μJ, the NIR pulses exiting the fiber experience six bounces on the CMs, while only four bounces on the CMs are used before collecting measurements at higher $E_p$. Black lines in the bottom panels of (a) and (b) show the laser spectrum and autocorrelation trace of the laser featuring a 3.5 THz spectral width (FWHM) and 185 fs pulse duration.

## III. RESULTS AND DISCUSSION

The NIR spectra after the HC-PCF and CMs is shown as a function of the input pulse energy $E_p$ in Fig. 2(a). At $E_p$ = 0.92 μJ, the spectrum reaches a spectral bandwidth of 7.7 THz full-width at half-maximum (FWHM). As the launched NIR pulse energy is increased up to 4.16 μJ, the bandwidth also gradually increases up to 32 THz (FWHM). At $E_p \geq$ 1.39 μJ, concentration of optical energy is observed within spectral lobes at the edges of the spectrum. This uneven energy distribution enables more efficient THz generation at frequencies corresponding to the spectral separation of these lobes [41]. The second-harmonic autocorrelation traces obtained with a 30-μm-thick beta-barium borate (BBO) crystal are shown in Fig. 2(b). For an input pulse energy of 0.92 μJ and after 6 bounces off the CMs, the pulse duration obtained after deconvolution is measured to be 106 fs FWHM, assuming a Gaussian pulse shape. As the input pulse energy increases up to 2.77 μJ, the pulse duration decreases, reaching a minimum of 31 fs. At higher input pulse energies, spectral broadening extends beyond the operational bandwidth of the CMs, which requires us to use 4 bounces to achieve the minimum pulse duration. We measure a pulse duration of 40 fs at $E_p$ = 3.23 μJ, 34 fs at $E_p$ = 3.70 μJ and 32 fs at $E_p$ = 4.16 μJ using the same configuration.

*Contact author: wcui2@uottawa.ca
*Contact author: jean-michel.menard@uottawa.ca

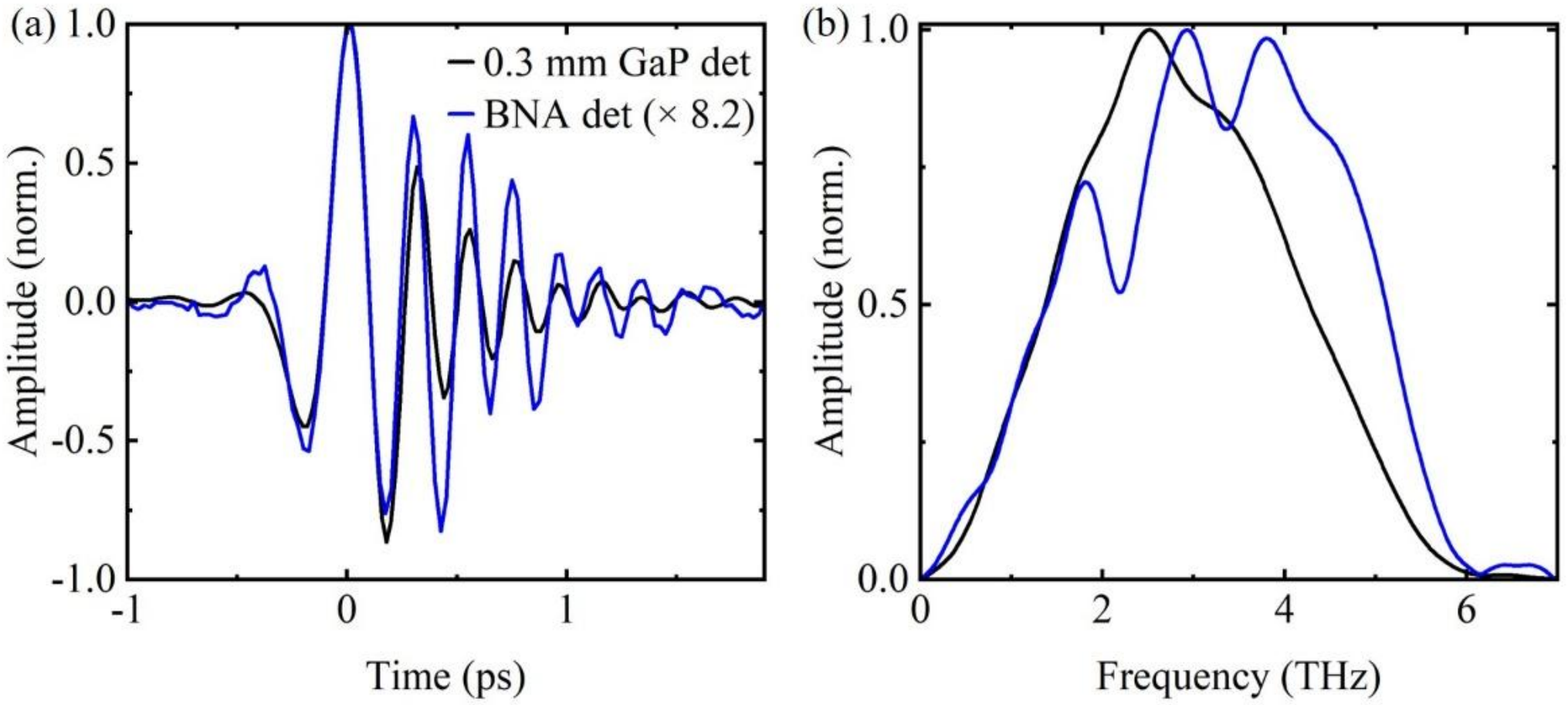

FIG 3. (a) THz transients and (b) spectral amplitudes measured using a 300-μm-thick GaP (black curve) and a 300 μm-thick free-standing BNA (blue curve) at $E_p$ = 2.31 μJ. The generation crystal is 1-mm-thick GaP crystal with a grating etched on the front surface [53,54]. In both plots, the THz waveforms and spectra are normalized to their respective maxima.

To evaluate the effectiveness of the BNA crystal for nonlinear THz detection, we compare the THz signals measured by electro-optic sampling with a 300 μm-thick free-standing BNA crystal and a reference 300 μm-thick GaP crystal. For these measurements, $E_p$ is set to 2.31 μJ, corresponding to a pulse duration of 35 fs. THz pulses centered around 2.6 THz and extending up to ~6 THz, are generated by tilted-pulse-front phase matching inside a 1-mm-thick GaP crystal, enabled by a phase grating etched directly onto the front surface of the crystal [53,54]. We monitor the THz temporal waveform and corresponding spectral amplitude obtained by the Fourier transform (Fig. 3). In the time domain, the peak signal detected with BNA reaches only 15% of that obtained with GaP, despite BNA exhibiting a larger second-order nonlinear coefficient [24]. Notably, substantial variations in signal amplitude are observed when the position of the gating pulse is shifted across the detection crystal. This reduction in EOS signal is attributed to BNA's intrinsic birefringence and spatial inhomogeneity. The birefringence can cause temporal walk-off and phase retardation between orthogonally polarized components of the gating pulse, which reduces the coherent electro-optic interaction. Meanwhile, crystal nonuniformity can introduce scattering, local optical-path variations and spatially varying birefringent retardance, which can distort both wavefront and polarization state of the gating beam, thereby reducing the coherent overlap required for efficient EOS. Further improvements in crystal fabrication, particularly toward more uniform single-crystal BNA, are therefore expected to enhance the EOS detection efficiency and improve the reproducibility of the response across the crystal aperture.

A frequency-domain analysis shows that the detected THz bandwidth using either BNA or GaP as the EO crystal extends to about 6 THz. This cutoff is not imposed by the detection crystals but instead by the limited bandwidth of the THz emission process. The measured spectrum with BNA includes phonon resonances at 2.2 and 3.3 THz, consistent with previous reports [31,55]. To generate higher-frequency components and access a broader spectral range, we replace the THz generation crystal with a 400-μm-thick BNA crystal. Figure 4 shows the resulting THz temporal waveforms and corresponding spectral amplitudes as we vary the input pulse energies $E_p$ in the HC-PCF-based pulse compressor. Increasing $E_p$ spectrally broadens and temporary compresses NIR pulses allowing access to a larger THz spectral window through broadband optical rectification and EOS sampling. At $E_p$ = 0.92 μJ, the THz time-domain signal appears as a multi-cycle pulse spanning over 2 ps. The corresponding spectrum extends up to 10.3 THz with most of the energy concentrated between 2.5 and 7.5 THz. At $E_p$ = 1.39 µJ, the time-domain signal exhibits higher frequency oscillations and a spectral component emerging at 11 THz. As we increase $E_p$ to 2.77 µJ, the THz spectrum efficiently covers low frequencies as well as the region between 7.5-12.5 THz. At $E_p$ = 3.23 µJ, new spectral components appear near 13.5 THz and 20.5 THz. Finally, at $E_p \geq 3.23$ µJ, additional THz components emerge between 17 and 25 THz. In the time domain, the maximum THz electric field measured at a pump pulse energy of $E_p$ = 3.70 µJ is approximately five times larger than that measured at $E_p$ = 0.92 µJ. In the spectral domain, the maximum spectral amplitude at $E_p$ = 4.16 µJ is approximately seven times larger than that at $E_p$ = 0.92 µJ, with the spectral peaks located at 10.8 THz and 5.3 THz, respectively. Since the present system provides broadband access across and beyond the "new THz gap", it offers a promising route to experimentally determine the high-frequency optical constants of BNA and other organic crystals, providing the data needed for future optimization studies.

*Contact author: wcui2@uottawa.ca
*Contact author: jean-michel.menard@uottawa.ca

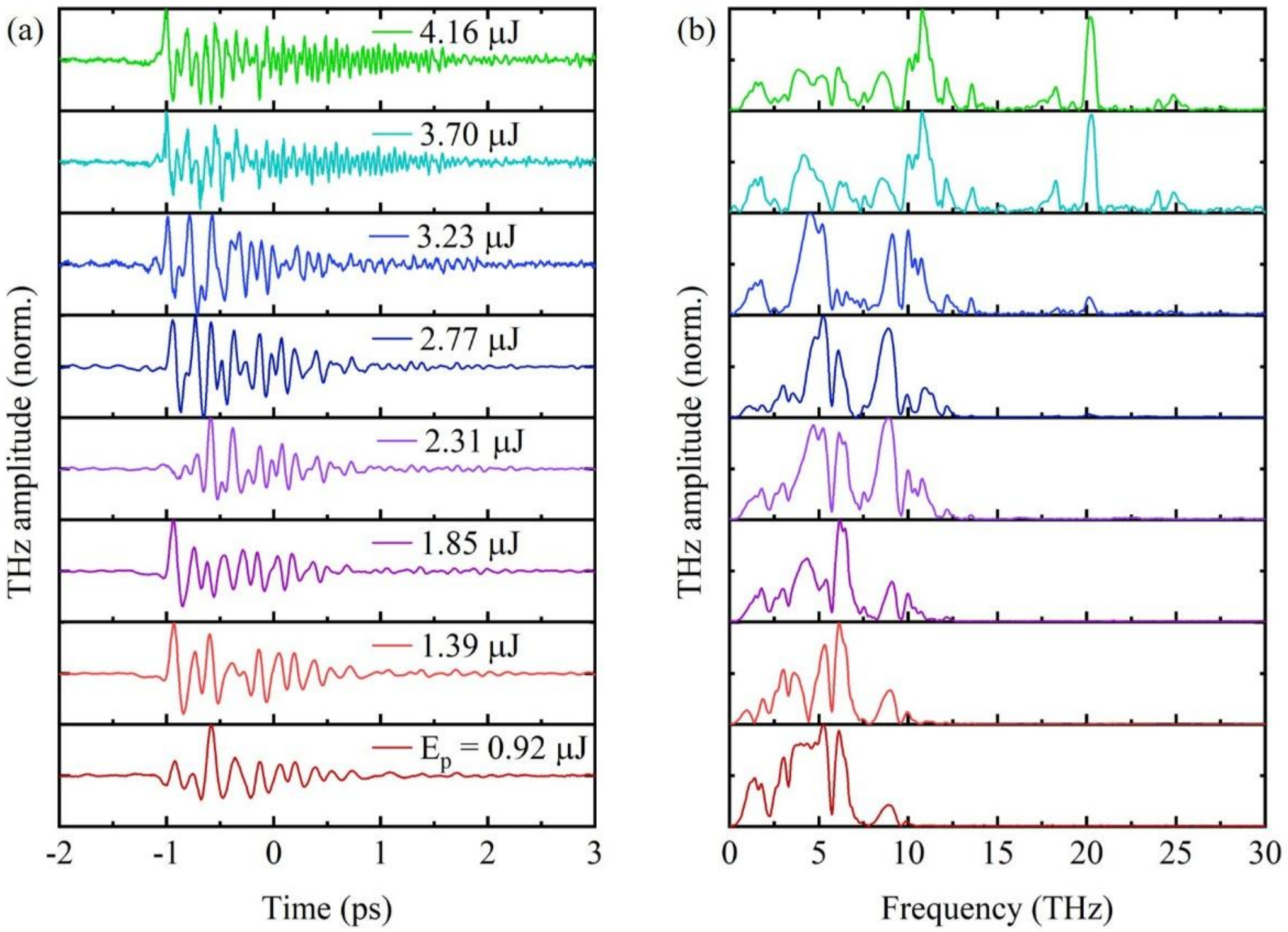

FIG 4. (a) Measured THz time-domain signal and (b) spectral amplitude obtained with $E_p$ ranging from 0.92 to 4.16 µJ. As $E_p$ increases, the time-domain signals display progressively faster oscillations, reflected in the appearance of higher-frequency components in the spectral domain.

We evaluate the sensitivity of our broadband THz-TDS system by characterizing the THz spectral intensity and noise floor at $E_p$ = 4.16 µJ. The results shown in Fig. 5 correspond to the average of six spectroscopy scans, with the light blue shaded area corresponding to the standard deviation. Each time-domain measurement is acquired over 8 minutes to cover a 5 ps scanning range with 10 fs resolution. The spectral noise floor, acquired under identical conditions with the THz beam blocked, is fitted to the model $A(1/f + B)$, which accounts for $1/f$ pink noise, where $f$ is the frequency and $A$ and $B$ are fitting parameters [1,53] . We observe a broad signal covering the window between 0.7 and 25.2 THz and featuring a maximum dynamic range (DR) of 55 dB at 11 THz. Notably, we find a DR generally approaching three orders of magnitude in the hard-to-access region between 3 and 13 THz. Phonon-related absorption dips are observed near 2.3, 3.3, 5.7, 7.3–7.9, 9.5, and 11.9 THz, consistent with the measured and simulated THz phonons [28]. In our system, these features arise from THz absorption in the generation and detection crystals. Despite these absorptions, the dynamic range remains robust and falls below 1000 only in limited regions. Previous experiments relying on two-beam difference-frequency generation in BNA [28,29] also demonstrated access to a similar spectral bandwidth. However, the generated spectrum with such a scheme is relatively narrowband and requires tunable, dual-wavelength NIR source to access the full spectral range. Here, our approach combining a Yb:KGW laser source and a single-ring HC-PCF provides access to a broadband THz region in a single measurement.

*Contact author: wcui2@uottawa.ca
*Contact author: jean-michel.menard@uottawa.ca

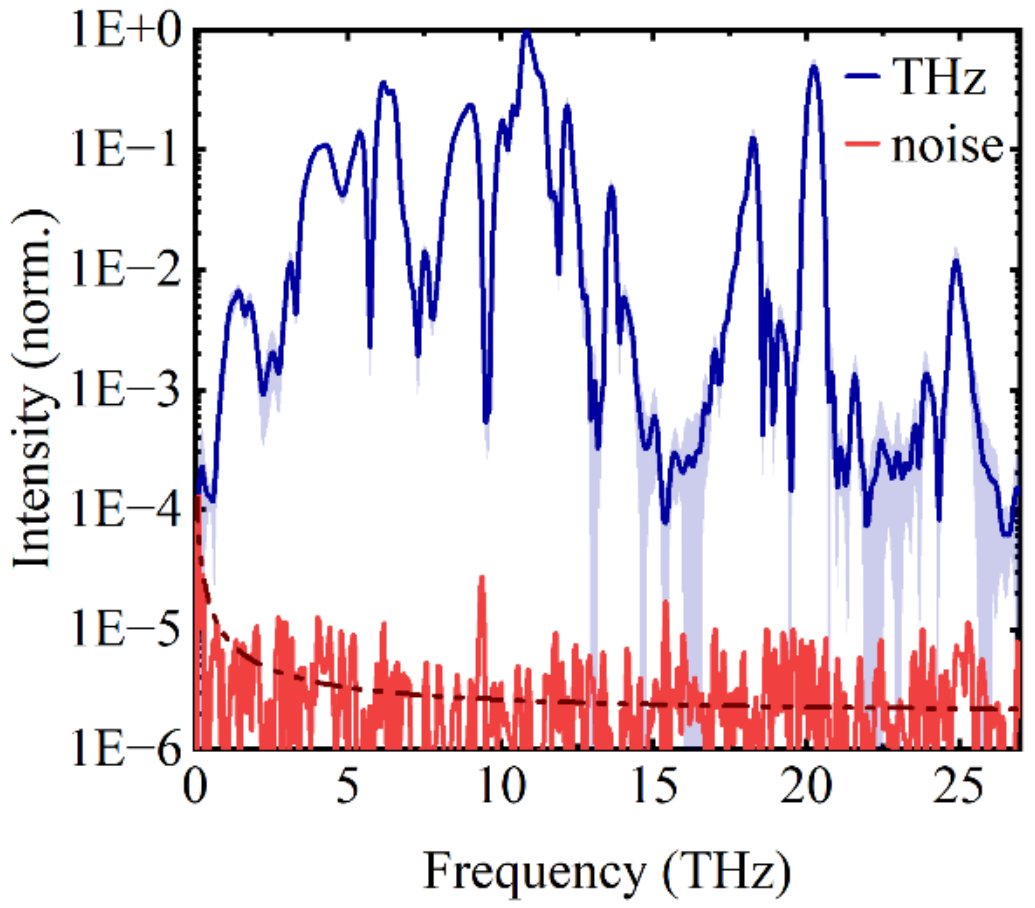

FIG 5. (a) Averaged THz spectral intensity (blue curve) with standard deviation (light blue shaded area), obtained by Fourier transforming individual time-domain scans and averaging in the spectral domain. The noise floor (red curve) was measured with the THz beam blocked. The dashed line is the noise floor fitted to the model $A(1/f+B)$, where f is the frequency, while $A$ and $B$ are fitting parameters [1,53].

## V. CONCLUSIONS

We present a THz-TDS system that uses a gas-filled single-ring HC-PCF to compress NIR pulses centered at 1.03 µm, enabling the generation and detection of broadband THz transients in BNA crystals. This approach provides access to a frequency window between 0.3 THz and 25.2 THz, unlocking new possibilities for spectroscopy applications. Interestingly, a significant portion of the generated THz energy falls between 3 and 13 THz, effectively addressing the "new THz gap," which has been challenging to access with conventional THz-TDS configurations. Additionally, our system leverages flourishing Yb:KGW laser technology, offering compact and cost-effective alternative to configurations relying on a dual-wavelength excitation for THz generation. Our results demonstrate a promising platform for sensitive, broadband THz spectroscopy, with potential applications in ultrafast carrier dynamics, vibrational spectroscopy of complex materials, and emerging fields such as chemical sensing, biomedical diagnostics, and high-speed optoelectronics.

## ACKNOWLEDGMENTS

J.-M.M. acknowledges funding from the Natural Sciences and Engineering Research Council of Canada (Grant No. RGPIN-2023-05365) and the Canada Foundation for Innovation (Project No. 35269). This work is also supported by the National Research Council of Canada via the High Throughput and Secure Networks Challenge Program (HTSN 254) and the Joint Centre for Extreme Photonics.

*Contact author: wcui2@uottawa.ca
*Contact author: jean-michel.menard@uottawa.ca

*Contact author: wcui2@uottawa.ca
*Contact author: jean-michel.menard@uottawa.ca

*Contact author: wcui2@uottawa.ca
*Contact author: jean-michel.menard@uottawa.ca

*Contact author: wcui2@uottawa.ca
*Contact author: jean-michel.menard@uottawa.ca